**Isotropic orbital magnetic moments in magnetically anisotropic $SrRuO_3$ films**

Yuki K. Wakabayashi,[1,*] Masaki Kobayashi,[2,3] Yukiharu Takeda,[4] Miho Kitamura,[5] Takahito Takeda,[3] Ryo Okano,[3] Yoshiharu Krockenberger,[1] Yoshitaka Taniyasu,[1] and Hideki Yamamoto[1]

[1]*NTT Basic Research Laboratories, NTT Corporation, Atsugi, Kanagawa 243-0198, Japan*
[2]*Center for Spintronics Research Network, The University of Tokyo, 7-3-1 Hongo, Bunkyo-ku, Tokyo 113-8656, Japan*
[3]*Department of Electrical Engineering and Information Systems, The University of Tokyo, Bunkyo, Tokyo 113-8656, Japan*
[4]*Materials Sciences Research Center, Japan Atomic Energy Agency, Sayo-gun, Hyogo 679-5148, Japan*
[5]*Photon Factory, Institute of Materials Structure Science, High Energy Accelerator Research Organization, Tsukuba, Ibaraki, 305-0801, Japan*

[*]Corresponding author: yuuki.wakabayashi.we@hco.ntt.co.jp

Abstract

Epitaxially strained $SrRuO_3$ films have been a model system for understanding the magnetic anisotropy in metallic oxides. In this paper, we investigate the anisotropy of the Ru 4*d* and O 2*p* electronic structure and magnetic properties using high-quality epitaxially strained (compressive and tensile) $SrRuO_3$ films grown by machine-learning-assisted molecular beam epitaxy. The element-specific magnetic properties and the hybridization between the Ru 4*d* and O 2*p* orbitals were characterized by Ru $M_{2,3}$-edge and O *K*-edge soft X-ray absorption spectroscopy and X-ray magnetic circular dichroism measurements. The magnetization curves for the Ru 4*d* and O 2*p* magnetic moments are identical, irrespective of the strain type, indicating the strong magnetic coupling between the Ru and O ions. The electronic structure and the orbital magnetic moment relative to the spin magnetic moment are isotropic despite the perpendicular and in-plane magnetic anisotropy in the compressive-strained and tensile-strained $SrRuO_3$ films; *i.e.*, the orbital magnetic moments have a negligibly small contribution to the magnetic anisotropy. This result contradicts Bruno's model, where magnetic anisotropy arises from the difference in the orbital magnetic moment between the perpendicular and in-plane directions. Contributions of strain-induced electric quadrupole moments to the magnetic anisotropy are discussed, too.

## I. INTRODUCTION

The itinerant 4*d* ferromagnetic perovskite $SrRuO_3$ (SRO) (bulk Curie temperature ($T_C$) = 165 K) is one of the most extensively studied ferromagnetic oxides because of the unique nature of its ferromagnetism, metallicity, chemical stability, and compatibility with other perovskite-structured oxides [1–22]. Epitaxial strain in SRO films is linked to the physical properties through the strong coupling among lattices, electrons (charges, spins), and orbits [23-26]. Particularly, a compressively strained SRO film on $SrTiO_3$ (STO) (001) is the first oxide heterostructure in which perpendicular magnetic anisotropy was discovered [9,27]. Generating perpendicular magnetic anisotropy is advantageous for scalability and reducing power consumption in spintronic devices, and thus SRO-based, all-oxide spintronic devices, have been investigated [17,18,28].

SRO has been a model system for understanding magnetic anisotropy in metallic oxides [10]. It is known that compressively and tensile strained SRO films have a perpendicular and in-plane magnetic easy axis, respectively [10,26]. The perpendicular magnetic anisotropy in compressively strained SRO has been considered to arise from magneto-crystalline anisotropy caused by spin-orbit interactions because a large Ru orbital magnetic moment perpendicular to the compressive-strained SRO films (0.08-0.1 $\mu_B$/Ru) below $T_C$ has been reported [29-31]. According to Bruno's model [32], the magneto-crystalline anisotropy energy is proportional to the difference in the orbital magnetic moment between the perpendicular and in-plane directions, as confirmed by XMCD for several systems, including Co thin films sandwiched by Au(111) [33] and FePt [34]. Therefore, to evaluate the contribution of the orbital magnetic moments to the magnetic anisotropy and examine whether or not Bruno's model [32] holds for SRO, determination of the anisotropy of the orbital magnetic moment of SRO having epitaxial compressive and tensile strain is a vital issue. In addition, recently, polarized neutron diffraction experiments revealed an unexpectedly large magnetic moment of O ions, which contributes 30% of the total magnetization in the case of bulk SRO [35], and XMCD measurements revealed the substantial orbital magnetic moment of the O 2*p* states induced by the hybridization with the Ru 4*d* $t_{2g}$ states in the case of compressive-strained SRO [31]. These results highlight the importance of the anisotropy of the Ru 4*d* and O 2*p* orbital magnetic moments and the orbital hybridizations for understanding the magnetic anisotropy and electronic structures in SRO films.

In this study, we investigated the anisotropy of the Ru 4*d* and O 2*p* electronic structure and magnetic properties using high-quality compressive-strained and tensile-strained SRO films by soft X-ray absorption spectroscopy (XAS) and XMCD, which are highly sensitive to the local electronic structure and element-specific magnetic properties in magnetic materials [36-40]. Ru $M_{2,3}$-edge and O *K*-edge absorptions were used in XAS and XMCD measurements. To synthesize SRO with even fewer crystalline defects than those in SRO in other reports, we employed our recently developed machine-learning-assisted molecular beam epitaxy (ML-MBE) [3,41,42]. We found that the electronic structure and the orbital magnetic moment relative to the spin magnetic moment are isotropic despite the perpendicular and in-plane magnetic anisotropy in the compressive and tensile-strained SRO films. This result contradicts Bruno's model. Furthermore, in both films, the magnetization curves of the Ru 4*d* and O 2*p* magnetic moments are identical, indicating the strong magnetic coupling of the Ru and O ions.

## II. EXPERIMENT

We grew high-quality epitaxial SRO films [Fig. 1(a)] with a thickness of 60 nm in a custom-designed molecular beam epitaxy (MBE) system with multiple e-beam evaporators for Sr and Ru [3,41,43]. To impose in-plane compressive and tensile epitaxial strains on SRO films, we used $SrTiO_3$ (STO) (001), $DyScO_3$ (DSO) (110), and $TbScO_3$ (TSO) (110) substrates. Rare-earth (*RE*) scandates $REScO_3$ have the $GdFeO_3$ structure, which is a distorted perovskite structure with the (110) face corresponding to the pseudocubic (001) face. The growth parameters were optimized by Bayesian optimization, a machine learning technique for parameter optimization [41,42,44,45], with which we achieved high residual resistivity ratios (RRRs) of 51, 72, and 20 for the SRO films on the STO, DSO, and TSO substrates, respectively. All SRO films were prepared under the same growth conditions as in Ref. [26], which also reported the electrical properties and crystallographic analyses of the films. The $T_C$ (152 K on STO, 169 K on DSO, and 164 K on TSO) and RRR values of the series of films are higher than or comparable to the highest reported values for SRO on each substrate [26], confirming the high crystalline quality of the films. All the films were grown coherently, and the in-plane strains of the SRO films on the STO, DSO, and TSO substrates were -0.7, 0.3, and 0.8%, respectively [26].

The samples were transferred in air to the helical undulator beamline BL23SU of SPring-8 [46-50] to perform the XAS and XMCD measurements. The monochromator resolution $E/\Delta E$ was about 10,000. The beam spot size was 200 × 100 $\mu m^2$ [49]. For the XMCD measurements, absorption spectra for circularly polarized X rays with the photon helicity parallel ($\mu^+$) and antiparallel ($\mu^-$) to the spin polarization were obtained by reversing the photon helicity at each photon energy $h\nu$ and recorded in the total-electron-yield (TEY) mode. The $\mu^+$ and $\mu^-$ spectra at the Ru $M_{2,3}$ edges and O *K* edge were taken for both positive and negative applied magnetic fields and averaged to eliminate spurious dichroic signals. The sample temperature was 5.5 or 6.5 K. For estimation of the integrated values of the XAS spectra at the Ru $M_{2,3}$ edges, hyperbolic tangent functions were subtracted from the spectra as background. To measure the perpendicular and in-plane magnetic moments, the angles of external magnetic fields and incident X rays measured from the sample surfaces were set to $\theta$ = 90° and 30°, respectively [Figs. 1(b) and 1(c)].

## III. RESULTS

Figure 2(a) shows the Ru $M_{2,3}$-edge XAS and XMCD spectra of the -0.7% compressive-strained SRO film on STO at 6.5 K with a magnetic field $\mu_0 H = 2$ T at $\theta$ = 90° (perpendicular). Here, $\mu^+$ and $\mu^-$ denote the absorption coefficients for the photon helicities parallel and antiparallel to the Ru 4*d* majority spin direction, respectively. In both the XAS and XMCD spectra, the absorption peaks from the Ru $3p_{3/2}$ and $3p_{1/2}$ core levels into the Ru 4*d* states are clearly observed at around 464 and 486 eV, respectively [51,52]. As in previous studies [30], the absorption peaks into Ru 5*s* states (477.5 eV) are only observed in the XAS spectra, indicating that the Ru 5*s* states have no spin polarization. Furthermore, the Ru $M_3$-edge XMCD peak position is lower than the Ru $M_3$-edge XAS peak position. Since the XMCD and XAS peaks come from the transitions to the unoccupied Ru 4*d* $t_{2g}$ and $e_g$ states [51], these different peak positions indicate that only the Ru 4*d* $t_{2g}$ states near the $E_F$ have spin polarization. These assignments are consistent with our previous Ru $M_{2,3}$ XMCD studies [31,51] and density functional theory (DFT) calculations [3], in which the half-metallic Ru 4*d* $t_{2g}$ states cross the $E_F$.

To clarify the unoccupied electronic states hybridized with the O 2*p* orbitals, we measured the O *K*-edge XAS and XMCD spectra. Figure 2(b) shows the O *K*-edge XAS and XMCD spectra of the -0.7% compressive-strained SRO film on STO at 6.5 K with a magnetic field of $\mu_0 H = 2$ T at $\theta$ = 90° (perpendicular). The O *K*-edge XAS spectra of 4*d* transition-metal oxides represent the transitions from the O 1*s* states to unoccupied transition-metal 4*d* and 5*s*/5*p* states, as well as those to the other conduction-band states via the hybridization with the O 2*p* states [53]. The XAS peaks at 529 and 530.4 eV have been attributed to absorptions into the coherent and incoherent parts of the Ru 4*d* $t_{2g}$ states, and the transitions to the Ru 4*d* $e_g$ states appear in the energy range of 532–534.5 eV [54,55]. The absorptions to the Sr 4*d* states and the Ru 5*s* states are observed in the 534.5–540 eV and 541–546 eV ranges [31,51,54,55], respectively. The energy difference between the Ru 4*d* $t_{2g}$ peak and the Ru 5*s* peak (~14.5 eV) positions is consistent with that in the Ru $M_{2,3}$-edge XAS and XMCD spectra (~14 eV) [Fig. 2(a)]. The intense coherent Ru 4*d* $t_{2g}$ peak indicates the itinerant nature (long lifetimes) of the quasiparticles in the hybridized O 2*p*-Ru 4*d* $t_{2g}$ states [31]. The long lifetimes are also evidenced by the quantum oscillations in the resistivity in SRO films with high RRRs over 20 [3,20,56]. The two-peak structure of the Ru 4*d* $e_g$ states is discernible in the spectra, which has not been seen in previously reported SRO films and bulk SRO [51,54,55,57], and is also consistent with our previous DFT calculations, in which the unoccupied Ru 4*d* $e_g$ states have the two-peak structure with peak separation of 1.5 eV [3]. From the O *K* XMCD spectrum [Fig. 2(b)], the substantial orbital magnetic moment of the O 2*p* states was observed, indicating that the half-metallic bands crossing the $E_F$ are formed by the Ru 4*d* $t_{2g}$ states hybridized with the O 2*p* states [3,31]. For the in-plane measurements ($\theta$ = 30°) [Figs. 2(c) and 2(d)], the Ru $M_{2,3}$ and O $K$-edge XAS and XMCD spectral shapes are almost the same as those for the perpendicular measurements ($\theta$ = 90°), but the Ru $M_{2,3}$ and O $K$-edge XMCD intensities are 1.8 times smaller than those for the perpendicular measurements. These results indicate that, like compressive-strained films reported previously [9,26,27,58], the -0.7% compressive-strained SRO film has perpendicular magnetic anisotropy.

The Ru $M_{2,3}$ and O $K$-edge spectra of the 0.3% and 0.8% tensile-strained SRO films on DSO (Fig. 3) and TSO (Fig. 4), respectively, have overall features similar to those of the -0.7% compressive-strained SRO film on STO. As in the case of the compressive-strained SRO film, the intense coherent Ru 4$d$ $t_{2g}$ peaks in O $K$-edge XAS spectra [Figs. 3(b), 3(d), 4(b), and 4(d)] indicate the long lifetimes of the quasiparticles in the hybridized O 2$p$-Ru 4$d$ $t_{2g}$ states. These results are consistent with the high metallic conductivity of the tensile-strained SRO films having the RRRs of 72 and 20 on DSO and TSO, respectively [26]. The Ru $M_{2,3}$ and O $K$-edge XMCD intensities for the 0.3% tensile-strained SRO film at $\theta$ = 30° are not much different from those at $\theta$ = 90° [Fig. 3]. In addition, the Ru $M_{2,3}$ and O $K$-edge XMCD intensities for the 0.8% tensile-strained SRO film at $\theta$ = 30° are 1.4 times larger than those at $\theta$ = 90° [Fig. 4]. These results indicate that the 0.3% tensile-strained SRO film does not have a large magnetic anisotropy and that the 0.8% tensile-strained SRO film has in-plane magnetic anisotropy [26,58,59]. Notably, the normalized Ru $M_{2,3}$-edge and O $K$-edge XMCD spectra for both in-plane ($\theta$ = 30°) and perpendicular ($\theta$ = 90°) measurements of all the SRO films are identical (Fig. 5), indicating that the orbital magnetic moment relative to the spin magnetic moment $m_{\mathrm{orb}}/m_{\mathrm{spin}}$ and the electronic structure are isotropic despite the perpendicular and in-plane magnetic anisotropy in the -0.7% compressive-strained and 0.8% tensile-strained SRO films. This result contradicts Bruno's model [32], and the origin of the magnetic anisotropy of epitaxially strained SRO films is different from that of Co thin films sandwiched by Au(111) [33] and FePt [34].

We determined the orbital magnetic moment $m_{\mathrm{orb}}$ and the spin magnetic moment $m_{\mathrm{spin}}$ of the $Ru^{4+}$ 4$d$ states using the XMCD sum rules as follows [36-38]:

$$m_{\mathrm{orb}} = -\frac{4(10-n_{4d})}{3r}\int_{M_2+M_3}(\mu^+ - \mu^-)dE,$$

$$m_{\mathrm{spin}} + 7m_{\mathrm{T}} = -\frac{2(10-n_{4d})}{r}\left[\int_{M_3}(\mu^+ - \mu^-)dE - 2\int_{M_2}(\mu^+ - \mu^-)dE\right].$$

Here, $r = \int_{M_2+M_3}(\mu^+ + \mu^-)dE$, and $n_{4d}$ is the number of electrons in 4$d$ orbitals, which is assumed to be four. For ions in octahedral symmetry, the magnetic dipole moment $m_{\mathrm{T}}$ is a small number compared to $m_{\mathrm{spin}}$ [60]. Using the XMCD spectra with a magnetic field of $\mu_0 H$ = 2 T [Figs. 2(a) and (c), 3(a) and (c), and 4(a) and (c)], we estimated the $m_{\mathrm{spin}}$ and $m_{\mathrm{orb}}$ values of $Ru^{4+}$ ions (Table I). The total magnetic moment, $m_{\mathrm{total}} = m_{\mathrm{spin}} + m_{\mathrm{orb}}$ = 0.757 $\mu_{\mathrm{B}}$/Ru for the -0.7% compressive-strained SRO film on STO with $\theta$ =

90°, is smaller than the saturation magnetization measured by a superconducting quantum interference device (SQUID) (1.25 $\mu_{\mathrm{B}}$ per pseudocubic unit cell) [31]. This discrepancy should arise from the magnetization of the O $2p$ electrons, which was verified from the O $K$-edge XMCD spectrum [Fig. 2(b)]. Assuming that the magnetic moment of oxygen contributes 30% of the total magnetization, as observed in bulk SRO [35], the sum of the Ru and O ion's magnetizations estimated from the $m_{\mathrm{total}}$ (0.757 $\mu_{\mathrm{B}}$/Ru) is 1.08 $\mu_{\mathrm{B}}$ per pseudocubic unit cell, which is close to that measured by the SQUID. The $m_{\mathrm{orb}}/m_{\mathrm{spin}}$ values of the Ru 4$d$ states were calculated from the $m_{\mathrm{spin}}$ and $m_{\mathrm{orb}}$ values in Table I. The constant $m_{\mathrm{orb}}$ / $m_{\mathrm{spin}}$ value of 0.08, regardless of magnetic anisotropy and measurement configuration, indicates the isotropic orbital magnetic moments in the magnetically anisotropic SRO films. In addition, irrespective of strain type, the ratios of the magnetizations of the $Ru^{4+}$ 4$d$ states at $\theta$ = 90° to that at $\theta$ = 30° at 2 T agree with those of the O $K$-edge XMCD intensities, meaning that the Ru magnetic moments and the magnetic moment of O are strongly coupled.

Figures 6(a) and (b) shows XMCD-$H$ curves measured at the Ru $M_3$ edge and the O $K$ edge for the -0.7% compressive-strained SRO film on STO at 6.5 K with $\theta$ = 90° and $\theta$ = 30°. The vertical axis of the XMCD intensity at 2 T has been scaled so that it represents the total magnetic moment $m_{\mathrm{total}}$ of the Ru ions estimated from Fig. 2(a). At $\theta$ = 90°, the remanent magnetization ratio of 0.94, estimated by the $m_{\mathrm{total}}$ values at 0 and $\pm 2$ T, is near an ideal value, indicating the almost single-domain perpendicular magnetization of the compressive-strained SRO film. The normalized XMCD-$H$ curves measured at the Ru $M_3$ edge and the O $K$ edge are identical, which confirms the strong coupling of the magnetic moments of the Ru 4$d$ and O 2$p$ electrons. The strong coupling of the magnetic moments of the Ru 4$d$ and O 2$p$ electrons is also observed in the tensile-strained SRO films [Figs. 6(c)–(f)]. The tensile-strained SRO films show the in-plane magnetic anisotropy: the magnetic moments at 2 T with $\theta$ = 30° are larger than those with $\theta$ = 90° [Table I] [26,58,59]. Compared to the 0.3% tensile-strained SRO film on DSO, the 0.8% tensile-strained SRO film on TSO has a larger ratio of magnetization at $\theta$ = 30° to that at $\theta$ = 90° and is more magnetically anisotropic.

## IV. DISCUSSION

The magnetic anisotropy in epitaxially strained SRO films has been thought to be interpreted by Bruno's model [29,30,32], in which the magneto-crystalline anisotropy energy is proportional to the difference in the orbital magnetic moment between the perpendicular and in-plane directions. However, the isotropic orbital magnetic moments in epitaxially strained SRO films mean that the contribution of the orbital magnetic moments is negligibly small. A possible source of the magnetic anisotropy in SRO is an electrical quadrupole moment, $\langle Q_{zz} \rangle = \langle 3m^2 - l(l+1) \rangle$, induced by the orbital distortion through the epitaxial strain. Here, $m$ and $l$ are the magnetic quantum number and angular momentum, respectively. The $\langle Q_{zz}\rangle$ stabilizes the magneto-crystalline anisotropy energy $K$ by the following equation [61,62,63]:

$$K = -\frac{21\xi^2}{4\mu_B \Delta E_{ex}}\langle Q_{zz}\rangle.$$

Here, $\Delta E_{ex}$ is the exchange splitting of the Ru 4$d$ bands, and $\xi$ is the spin-orbit coupling constant. $K > 0$ and $K < 0$ correspond to perpendicular and in-plane magnetic easy axes, respectively, and thus a negative $\langle Q_{zz}\rangle$ value of Ru 4$d$ orbitals stabilizes the perpendicular magnetic anisotropy. Due to the large spin-orbit coupling constant $\xi$ of Ru 4$d$ orbitals [10] compared to 3$d$ transition metals, the contribution of electric quadrupole moments could be larger than other 3$d$ perpendicular magnetic thin films, such as FePt [34,64], $La_{1-x}Sr_xMnO_3$ [65], Fe/MgO [66], and $Mn_{3-\delta}Ga$ [63]. Under the compressive strain from a STO substrate, the energy of the Ru 4$d$ $t_{2g}$ $d_{xz}$ and $d_{yz}$ orbitals [$\langle Q_{zz}\rangle = -1$] would be lower than that of the Ru 4$d$ $t_{2g}$ $d_{xy}$ orbital [$\langle Q_{zz}\rangle = 2$] in a SRO film, resulting in preferential occupation in $d_{xz}$ and $d_{yz}$ orbitals and the perpendicular magnetic anisotropy [57,67]. For further clarification of the contribution of the electric quadrupole, systematic characterization of the $\langle Q_{zz}\rangle$ by X-ray magnetic linear dichroism (XMLD) for high-quality epitaxially strained SRO films remains as future work [63,65].

## V. CONCLUSIONS

We have systematically investigated the anisotropy of the orbital magnetic moments and the orbital hybridizations of the Ru 4$d$ and O 2$p$ states in high-quality epitaxially strained SRO films. We found the isotropic orbital magnetic moments and isotropic O 2$p$-Ru 4$d$ $t_{2g}$ hybridizations despite the perpendicular and in-plane magnetic anisotropy in compressive and tensile-strained SRO films. The Ru 4$d$ and O 2$p$ magnetic moments are strongly coupled. The isotropic orbital magnetic moments reveal the negligibly small contribution of the orbital magnetic moments in the magnetic anisotropy. Although further detailed investigations are necessary to clarify the contribution of the electric quadrupole, which will affect the magnetic anisotropy of epitaxial SRO films through the large spin-orbit coupling of the Ru 4$d$ orbitals, our results provide important insights into the magnetic anisotropy and orbital hybridizations in epitaxially strained SRO films.

## ACKNOWLEDGMENTS

This work was partially supported by the Spintronics Research Network of Japan (Spin-RNJ). This work was performed under the Shared Use Program of Japan Atomic Energy Agency (JAEA) Facilities (Proposal No. 2021B-E25) supported by JAEA Advanced Characterization Nanotechnology Platform as a program of the ”Nanotechnology Platform“ of the Ministry of Education, Culture, Sports, Science and Technology (MEXT) (Proposal No. JPMXP09A21AE0045). The experiment at SPring-8 was approved by the Japan Synchrotron Radiation Research Institute (JASRI) Proposal Review Committee (Proposal No. 2021B3841).

## AUTHORS' CONTRIBUTIONS

Y.K.W. and M.K. conceived the idea, designed the experiments, and directed and supervised the project. M.K. and Y.K.W. planned the synchrotron experiments. Y.K.W. and Y.K. grew the samples. Y.K.W. carried out the sample characterizations. Y.K.W., M.Ko., M.Ki., Y.Tak., T.T., and R.O. carried out the XMCD measurements. Y.K.W. and M.K. analyzed and interpreted the Data. Y.K.W. wrote the paper with input from all authors.

## DATA AVAILABILITY

The data that support the findings of this study are available from the corresponding authors upon reasonable request.

**Figures and figure captions**

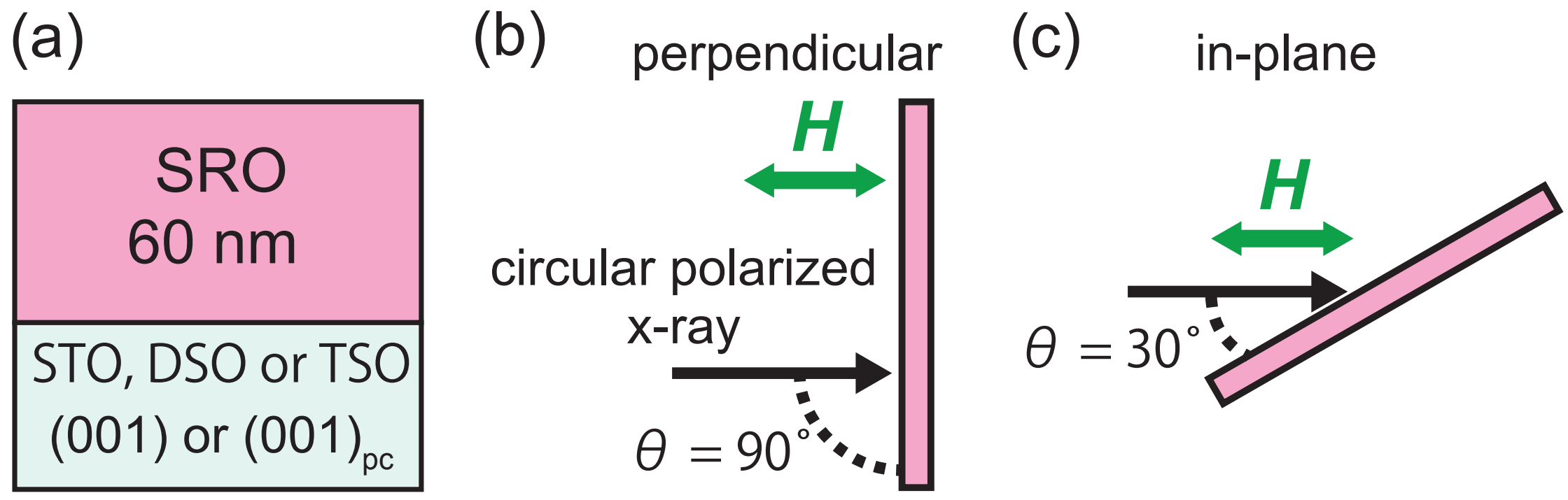

FIG. 1. Schematic illustrations of (a) the sample structure and measurement configurations for the (b) perpendicular and (c) in-plane XAS and XMCD measurements. In (a), $(001)_{pc}$ represents the pseudocubic (001) direction.

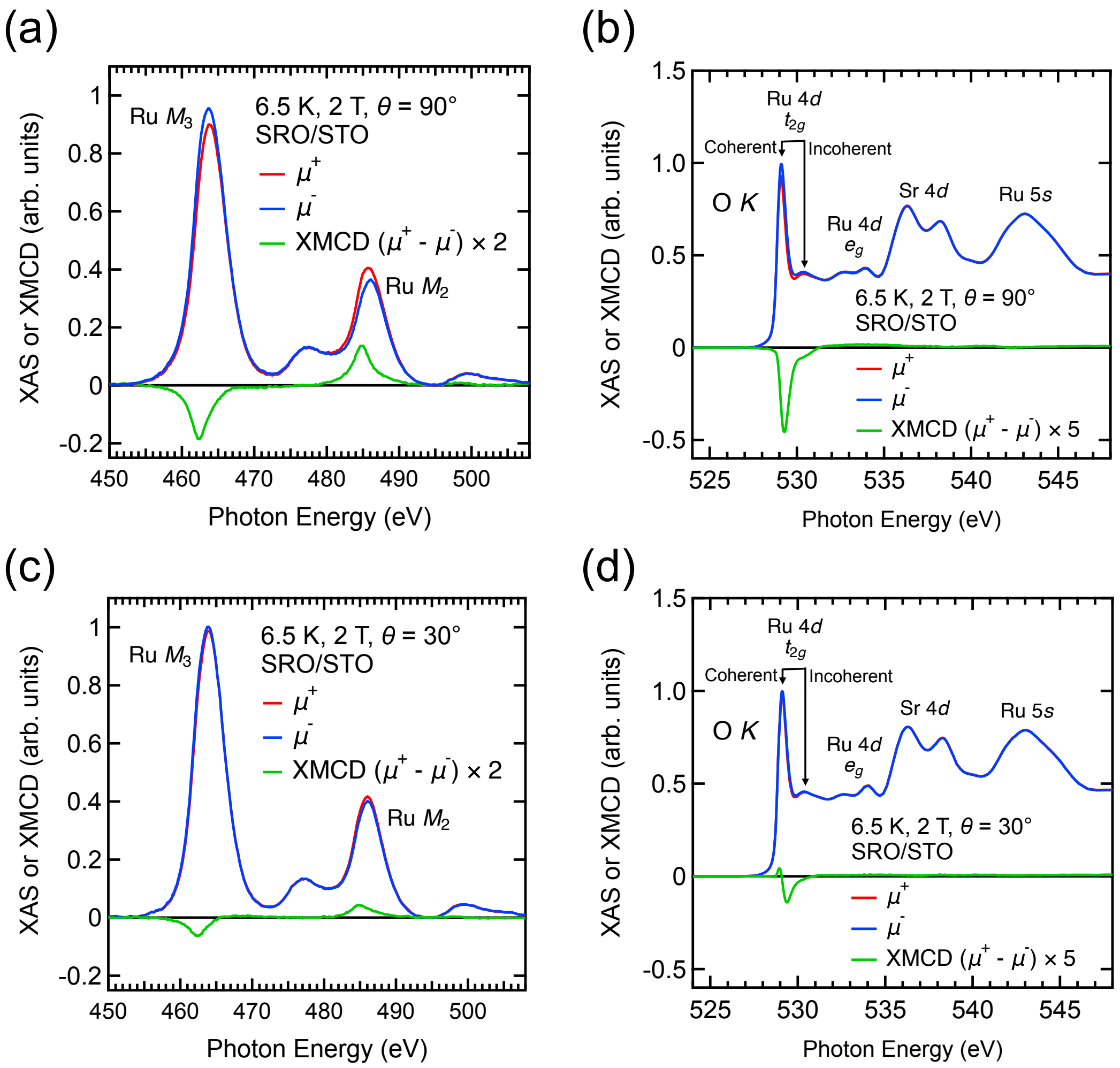

FIG. 2. (a), (c) Ru $M_{2,3}$-edge XAS and XMCD spectra for the -0.7% compressive-strained SRO film on STO at 6.5 K with a magnetic field $\mu_0 H = 2$ T at (a) $\theta = 90°$ and (c) $\theta = 30°$. (b), (d) O $K$-edge XAS and XMCD spectra for the -0.7% compressive-strained SRO film on STO at 6.5 K with a magnetic field $\mu_0 H = 2$ T at (b) $\theta = 90°$ and (d) $\theta = 30°$.

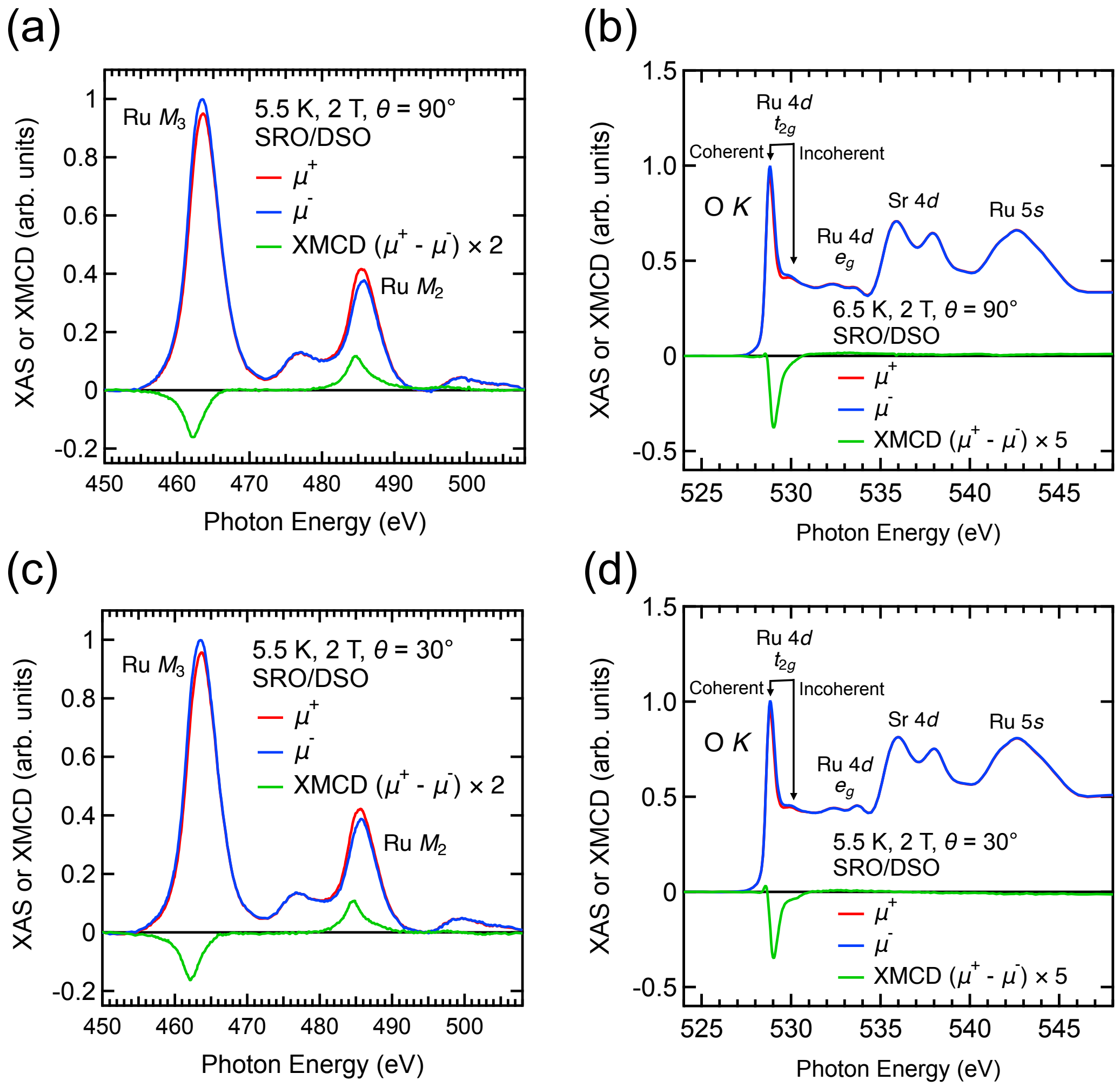

FIG. 3. (a), (c) Ru $M_{2,3}$-edge XAS and XMCD spectra for the 0.3% tensile-strained SRO film on DSO at 5.5 K with a magnetic field $\mu_0 H = 2$ T at (a) $\theta = 90°$ and (c) $\theta = 30°$. (b), (d) O $K$-edge XAS and XMCD spectra for the 0.3% tensile-strained SRO film on DSO at 5.5 K with a magnetic field $\mu_0 H = 2$ T at (b) $\theta = 90°$ and (d) $\theta = 30°$.

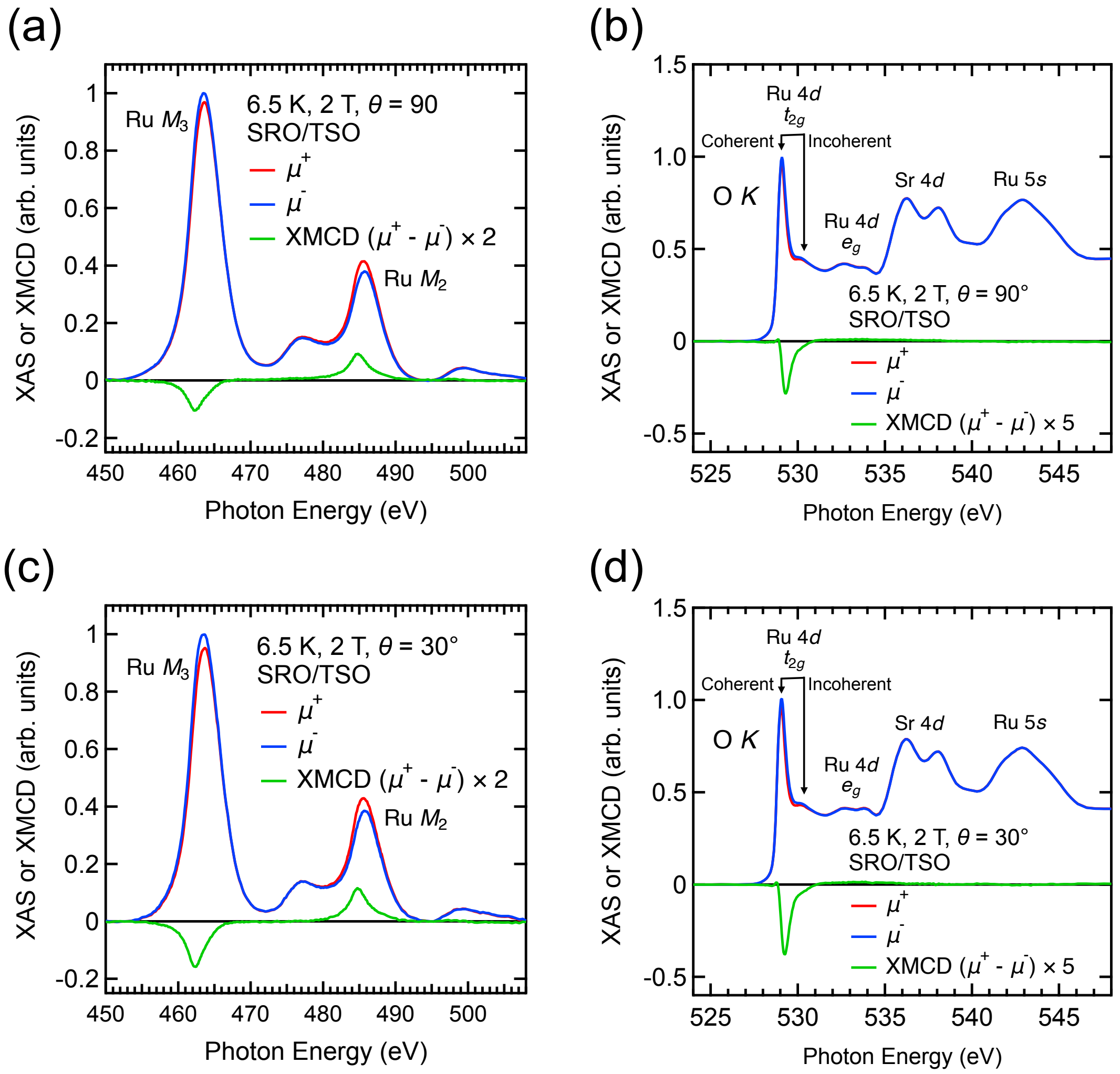

FIG. 4. (a), (c) Ru $M_{2,3}$-edge XAS and XMCD spectra for the 0.8% tensile-strained SRO film on TSO at 6.5 K with a magnetic field $\mu_0 H = 2$ T at (a) $\theta = 90°$ and (c) $\theta = 30°$. (b), (d) O $K$-edge XAS and XMCD spectra for the 0.8% tensile-strained SRO film on TSO at 6.5 K with a magnetic field $\mu_0 H = 2$ T at (b) $\theta = 90°$ and (d) $\theta = 30°$.

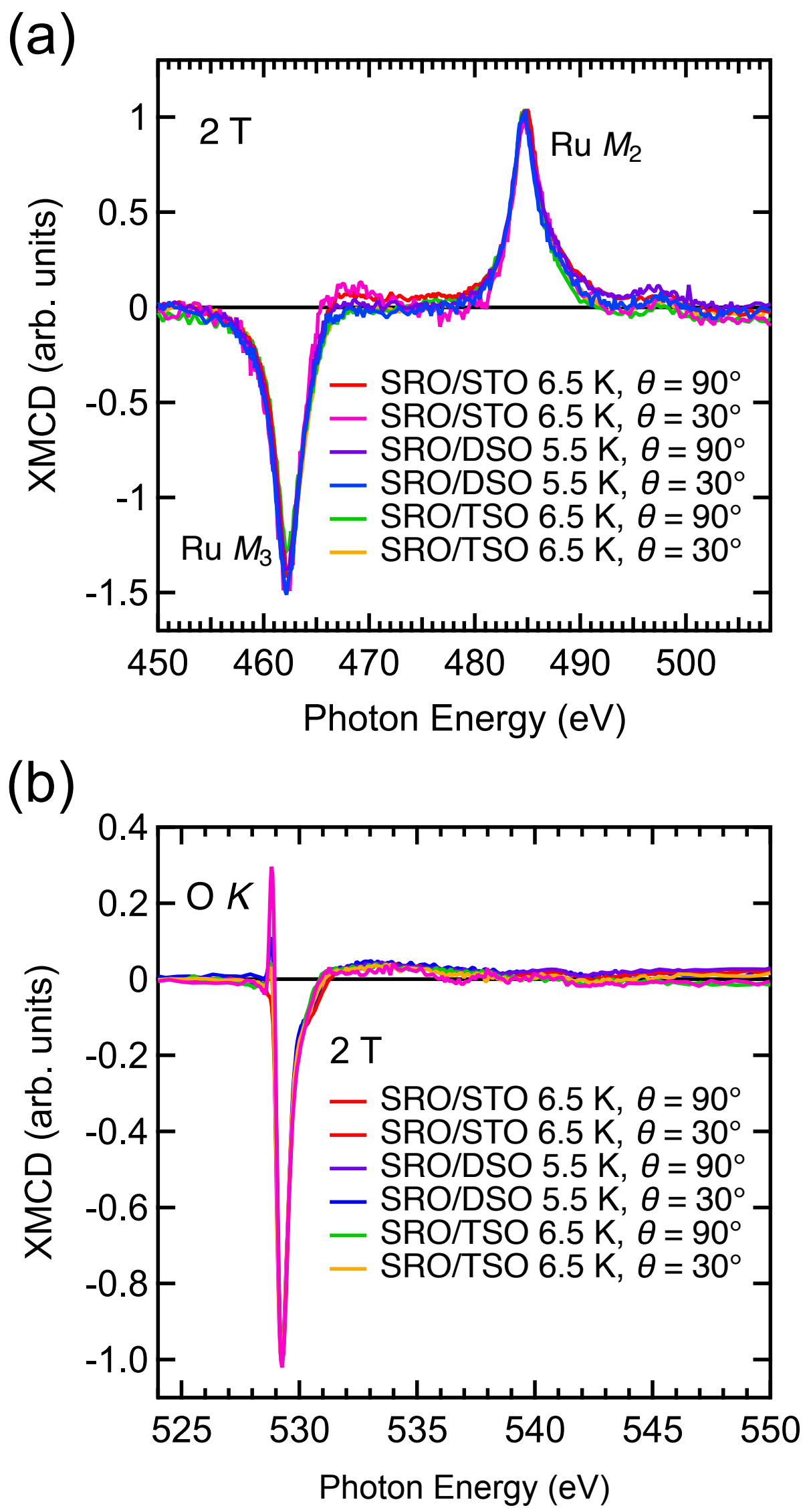

FIG. 5. (a) Ru $M_{2,3}$-edge spectra normalized at the Ru $M_2$-edge XMCD peak and (b) O $K$-edge XMCD spectra normalized at the O $K$-edge XMCD peak for the SRO films with a magnetic field $\mu_0 H = 2$ T at $\theta = 90°$ and $\theta = 30°$ at 5.5 or 6.5 K.

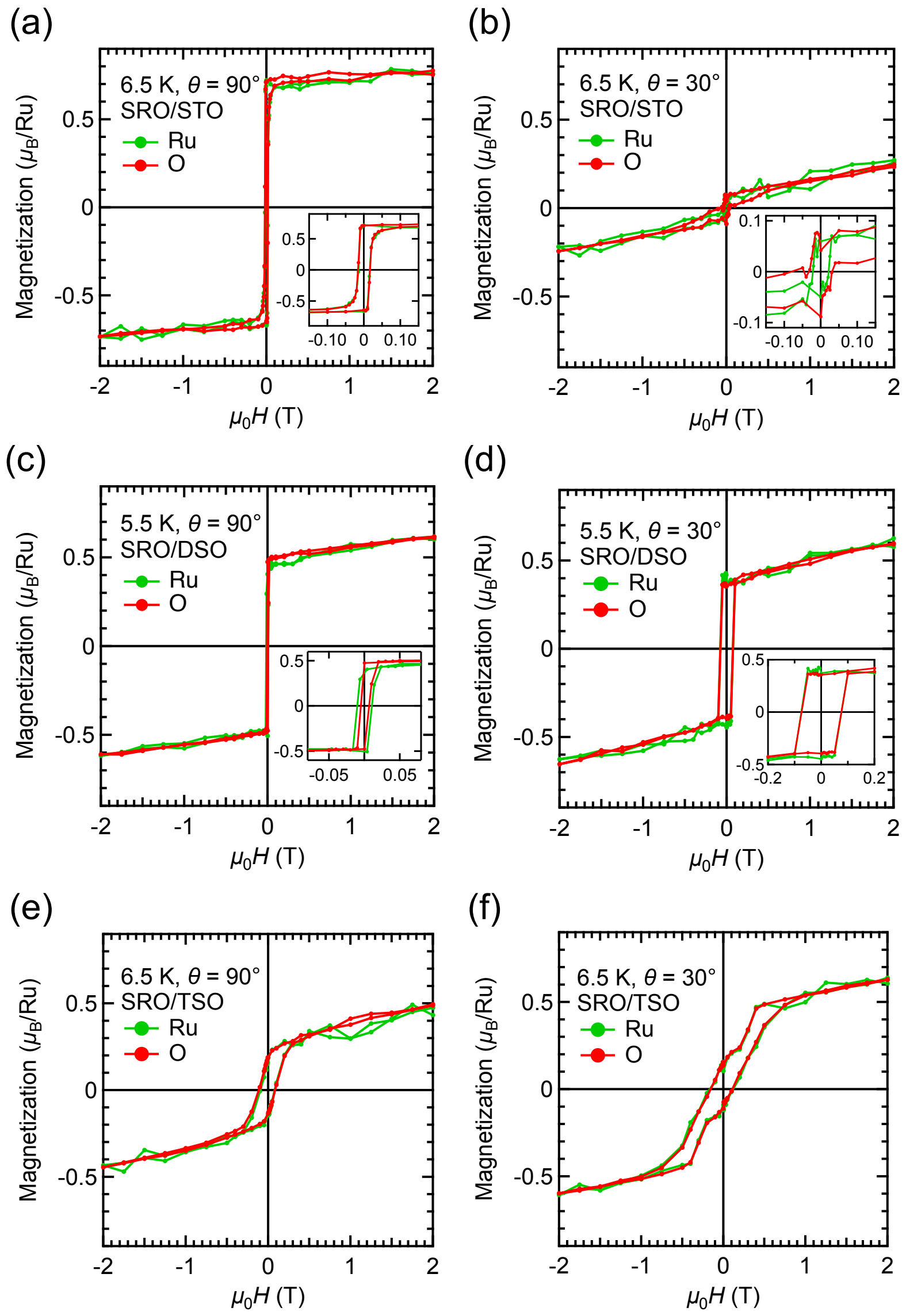

FIG. 6. XMCD-$H$ curves measured at the Ru $M_3$-edge and the O $K$-edge for (a), (b) the -0.7% compressive-strained SRO film on STO, (c), (d) the 0.3% tensile-strained SRO film on DSO, and (e), (f) the 0.8% tensile-strained SRO film on TSO at 5.5 or 6.5 K with (a),(c),(e) $\theta$ = 90° and (b),(d),(f) $\theta$ = 30°.

TABLE I. $m_{spin}$, $m_{orb}$, and $m_{total}$ [$\mu_B$/Ru] values estimated from the XMCD sum rules.

| Sample | $m_{spin}$ [$\mu_B$/Ru] | $m_{orb}$ [$\mu_B$/Ru] | $m_{total}$ [$\mu_B$/Ru] |
|---|---|---|---|
| SRO/STO $\theta = 30°$ | 0.22 | 0.018 | 0.238 |
| SRO/STO $\theta = 90°$ | 0.70 | 0.057 | 0.757 |
| SRO/TSO $\theta = 30°$ | 0.58 | 0.046 | 0.626 |
| SRO/TSO $\theta = 90°$ | 0.57 | 0.047 | 0.617 |
| SRO/DSO $\theta = 30°$ | 0.56 | 0.046 | 0.606 |
| SRO/DSO $\theta = 90°$ | 0.40 | 0.033 | 0.433 |